# EMPHASIS: An Emotional Phoneme-based Acoustic Model for Speech Synthesis System

*Hao Li, Yongguo Kang, Zhenyu Wang*

Baidu Speech Department
Baidu Inc. Baidu Technology Park, Beijing, 100193, China
{lihao20, kangyongguo, wangzhenyu06}@baidu.com

## Abstract

We present EMPHASIS, an emotional phoneme-based acoustic model for speech synthesis system. EMPHASIS includes a phoneme duration prediction model and an acoustic parameter prediction model. It uses a CBHG-based regression network to model the dependencies between linguistic features and acoustic features. We modify the input and output layer structures of the network to improve the performance. For the linguistic features, we apply a feature grouping strategy to enhance emotional and prosodic features. The acoustic parameters are designed to be suitable for the regression task and waveform reconstruction. EMPHASIS can synthesize speech in real-time and generate expressive interrogative and exclamatory speech with high audio quality. EMPHASIS is designed to be a multi-lingual model and can synthesize Mandarin-English speech for now. In the experiment of emotional speech synthesis, it achieves better subjective results than other real-time speech synthesis systems.

## 1. Introduction

Emotional speech synthesis has been a hotspot for decades in the research on statistical parametric speech synthesis (SPSS）[1]. In the last few years, artificial neural network（ANN）has become the most popular method for SPSS. The networks can offer efficient representation of the complex dependencies between linguistic and acoustic features, such as recurrent neural networks (RNN) with long-short-term-memory unit (LSTM) [2], [3] and gated-recurrent-unit (GRU) [4], convolutional neural networks and their variants [5], [6]. These models have provided elegant ways to model speech data and successfully applied to acoustic modeling for SPSS. ANN-based method is also being applied for emotional speech synthesis and achieves natural emotional expressions [7], [8].

Generally speaking, the ANN-based acoustic model of SPSS has two kinds of structures, the cascade model structure and the end-to-end structure. The cascade model structure, such as LSTM-based network [2], [3] and Deep Voice [5] etc., include a duration model and an acoustic parameter model. The high level linguistic features are firstly fed to the duration model to get the duration of each input vector, and then up-sampled to frame-level sequences for acoustic parameter prediction. This kind of model structure is generally stable because it has reasonable durations and strict frame-to-frame mapping between linguistic features and acoustic parameters. It is able to generate speech within acceptable time. However, systems using cascade model structure usually have over-smooth effect, which will lead to fuzzy speech. The end-to-end structure, such as Tacotron [9] and Char2Wav [10], etc., need no duration model. These models are based on sequence-to-sequence mapping [11] with attention mechanism [12]. By using this structure, one need no phoneme segmentations to train the network, which can greatly reduce labor work. End-to-end model structure is relatively difficult to train and more time-consuming under current stage.

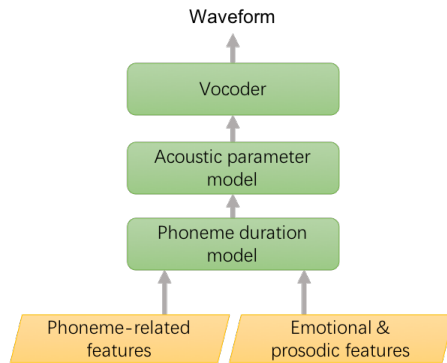

Figure 1: *Framework of cascade acoustic model for emotional statistical parametric speech synthesis.*

In this paper, we present a new acoustic model for SPSS: EMPHASIS, short for the emotional phoneme-based acoustic model for speech synthesis system. EMPHASIS uses cascade model structure, so it is stable and almost has no failure case. We use a network structure based on stacked 1-D convolution banks, highway layers and bi-directional GRU layers (CBHG) [9], [13]. In order to reduce the over-smooth effect, we adjust the output layer to be suitable for generating acoustic parameters. In order to have better emotional effect, we apply a feature grouping strategy in the input layer to extract emotion-related features and other linguistic features independently. EMPHASIS is designed to be a multi-lingual model and can synthesize Mandarin-English speech for now. It can synthesize high-quality emotional speech, including speech with interrogative expression and exclamatory expression.

This paper is organized as follows. Section 2 describes the framework of our system and the details of the duration model and acoustic parameter model; Section 3 gives the experiments and results of subjective evaluation; Section 4 is the conclusion.

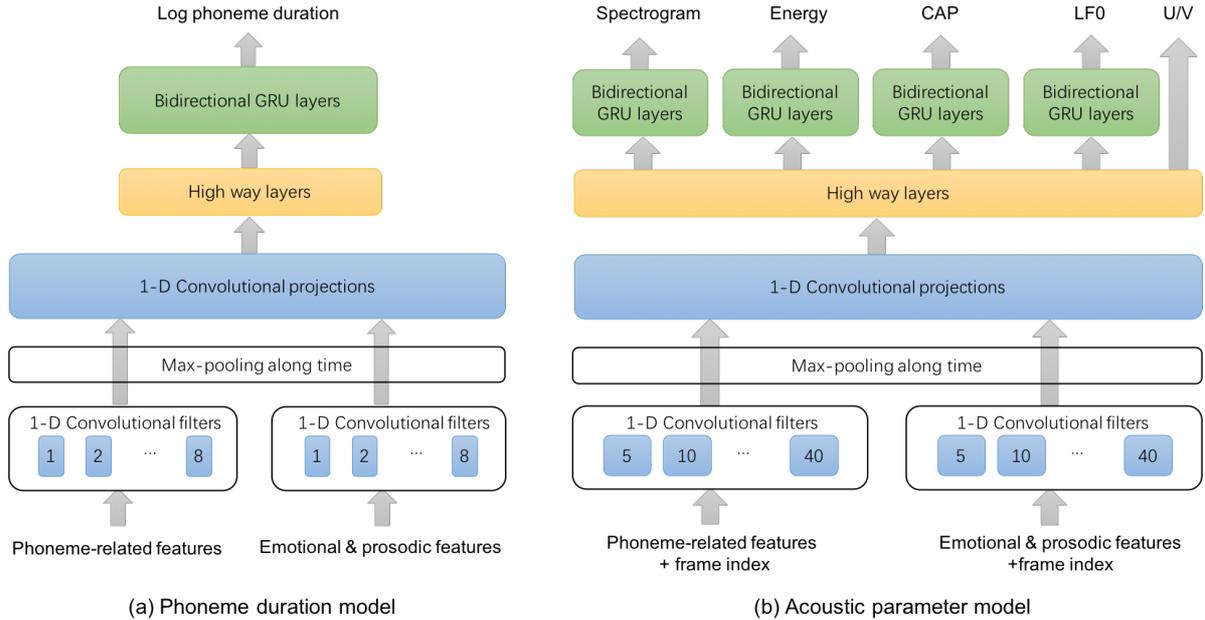

Figure 2: *Model architecture of EMPHASIS.*

## 2. Model architecture

The framework of EMPHASIS is shown in Figure 1. The phoneme-level linguistic features are first fed to a duration model, and then up-sampled to frame-level by the predicted phoneme durations, these features are further fed to the acoustic parameter model. Finally, a vocoder is used to reconstruct the waveform from the acoustic parameters. The duration model and acoustic parameter prediction model are based on the CBHG [9], [13] structure and a linguistic feature grouping strategy.

### 2.1. CBHG prototype

CBHG is a powerful module for extracting representation from sequences, which has been adopted as building block of Tacotron [9], [6], a successful end-to-end speech synthesis model.

In CBHG, the input sequence is first convolved with several 1-D convolutional filters (convolution banks). These filters explicitly model local and contextual linguistic information. The convolution outputs are stacked together and max pooled along time to increase local invariances. The processed sequence is fed to a few fixed-width 1D convolutions, which are used to project the stacked output into a lower dimension. The output is then fed into a multi-layer highway network to extract high-level features. Stacked bidirectional GRU layers are used on the top of the network to extract sequential features from both forward and backward context.

Like Tacotron, we use non-causal convolutions, batch normalization for the projection convolutional layers and stride 1 max pooling, which are proven to be necessary. We do not use any residual connection and batch normalization for convolution banks, which are not helpful according to our tests.

### 2.2. Linguistic feature grouping

The linguistic features of a text-to-speech (TTS) system usually include phoneme identification, tone (for tonal language, i.e. Mandarin), stress (for non-tonal language, i.e. English), prosodic structure, syntactical structure and type of emotion, etc. These features are converted into one-hot representation before fed to the network. The phoneme-related features, including phoneme identification, tone and stress, are very strong features for determining the pattern of durations and acoustic features, while the emotional & prosodic features, including prosodic structure, syntactical structure and type of emotion, are relatively week features. Therefore, during the training, the networks are very likely to overfit the phoneme-related features and, to some degree, ignore the emotional & prosodic features. Therefore, we separate the linguistic features into phoneme-related group and emotional & prosodic group, so that we can enhance the latter by certain structure of the network. In EMPHASIS, we use two groups of convolution filters to accept these two groups of features, respectively, and make sure these two kinds of features do not interfuse at low level of the network. In our tests, this structure achieves better emotional speech than the structure with only one filter group.

### 2.3. Phoneme duration prediction model

The phoneme duration prediction model is shown in Figure 2(a). The main difference between this model and the CBHG prototype is that it uses two groups of 1-D convolutional filters, as discussed above. Each group of convolutional filters contains 8 filters of width 1, 2, …, 8. The continuous-valued durations of phonemes are transformed into logarithmic domain to make them more Gaussian. When we use root mean square error (RMSE) loss to train the network, the logarithmic transformation increases the accuracy and variance of the predicted duration.

**2.4. Acoustic parameter prediction model**

The acoustic parameter prediction model also uses two groups of 1-D convolutional filters. Each group contains 8 filters. Because the input sequences are frame-level, they are much longer than the original phoneme-level input sequences, therefore, the filter widths are set to 5, 10, 15, 20, …, 40, so that they can see wider contextual information.

Another difference between this model and the CBHG prototype is that it uses independent stacked bidirectional GRU layers for different acoustic parameters. As shown in Figure 2(b). Four groups of stacked GRU layers are used to generate spectrogram, energy, aperiodicity parameter and log fundamental frequency (LF0), respectively. Note that there is no GRU layer for the unvoiced/voiced (U/V) parameter, because it is relatively much easier to predict, and we use cross-entropy loss rather than RMSE loss for U/V because it has 0-1 distribution.

By separating the acoustic parameters using independent bidirectional GRU layers, the interferences between the parameters are reduced. This design can improve the audio quality, and more importantly, gives better energy and LF0 curves, which are highly correlated with human emotion.

**2.5. Acoustic parameters**

We use WORLD [14], [15] vocoder to reconstruct speech waveform from acoustic parameters. It is a high-quality speech analysis/synthesis system and one of the most popular vocoders for SPSS. The acoustic parameters include spectrogram, aperiodicity, and fundamental frequency. Instead of using the parameters directly, we use the following transformations to make them more flexible and easier to predict by a regression network.

- *Energy & Spectrogram*: We extract energy from each frame of waveform and remove the energy component from the original spectrogram, so that the energy can be explicitly described.
- *Aperiodicity*: We predict aperiodicity of center frequencies (CAP) instead of the spectral representation according to [14]. The aperiodicity of the spectral representation is obtained by linear interpolation from CAP.
- *LF0 & U/V*: The fundamental frequency is transformed into LF0 and U/V features, which is a common procedural.

# 3. Experiments and Discussions

The experimental dataset contains 14 hours' Mandarin speech and 3.5 hours' English speech of a female speaker. The text of the dataset contains 30% interrogative sentences and 10% exclamatory sentences, which the speaker read with natural emotion. The audio files are segmented into utterances and down-sampled to 16kHz. The linguistic information and phoneme segmentation are labeled manually. 5% of the data is used as validation set to prevent over-training.

The phoneme-related linguistic features include phoneme identity (specifically, one-hot representation of initials and finals of Mandarin Pinyin and phonemes of English）, tone of Mandarin syllable, stress of English syllable; the emotional & prosodic linguistic features include prosodic break, prosodic level, word segmentation of Mandarin, syntactic level, part-of-speech tag and type of emotion, etc.

Table 1: *Network configuration of EMPHASIS in experiment.*

| Hyper parameters | Duration model | Acoustic parameter model | |
|---|---|---|---|
| Convolutional bank widths | 1, 2, … ,8 | 5, 10, …, 40 | |
| Max pooling width | 2 | 10 | |
| High way layers | 1 | 2 | |
| GRU (#layer * #node * directions) | 2 * 32 * 2 | Spec. | 2 * 128 * 2 |
| | | Energy | 2 * 16 * 2 |
| | | CAP | 2 * 16 * 2 |
| | | LF0 | 2 * 32 * 2 |

**3.1. ABX preferences tests**

We compare EMPHASIS with bidirectional-LSTM-based model (Bi-LSTM), Tacotron and concatenative TTS system.

- *EMPHASIS*: The network configurations are shown in Table 1. The convolutional filters can extract both local and contextual information, so there is no need to use phoneme-level features of prior and following phonemes, each input vector contains only the information related to the current phoneme.
- *Tacotron*: Settings of Tacotron are from [9], with the Griffin-Lim reconstruction algorithm [16] to synthesize speech.
- *Bi-LSTM*: The structure of this system is stacked 2 fully connected (FC) layers and 2 bidirectional LSTM layers, as in [3]. It is a state-based model because the target of the duration model is hidden-Markov model (HMM) state durations. The node numbers of hidden layers of duration model and acoustic parameter model are 32 and 128, respectively, which are selected based on the performances on the validation set. It uses WORLD vocoder to reconstruct waveform.
- *Concatenative*: This is an HMM-based unit selection concatenative speech synthesis system.

We conducted several ABX preference tests between EMPHASIS and the other three systems. The test sentences include 10 declarative sentences, 10 interrogative sentences and 10 exclamatory sentences. Five of the sentences contain English words. (Examples of synthesized speech are available at https://ttsdemos.github.io.) Ten subjects, who are not professional researcher and all of whom use Mandarin as mother tongue and English as second language, participate in the tests. The results are shown in Figure 3.

**3.2. Discussions**

From the ABX preference scores in Figure 3. We can see that EMPHASIS is significantly better than the Bi-LSTM system. Considering that EMPHASIS has deeper network structure and more weights, this is a reasonable result. The Bi-LSTM system produced fuzzy speech due to the over-smoothness of the acoustic parameters, while this phenomenon was alleviated in our system. Note that the Bi-LSTM structure is very fast for generating speech, its real-time ratio is less than 0.1, while our system has 0.3~0.4 real-time ratio under the same hardware and software environment.

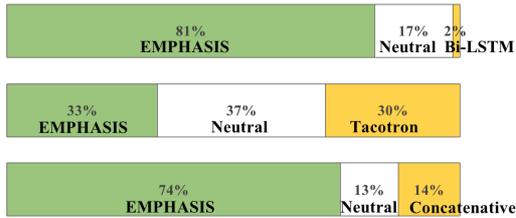

Figure 3：*ABX preference scores of EMPHASIS vs. Bi-LSTM, Tacotron, and Concatenative system*

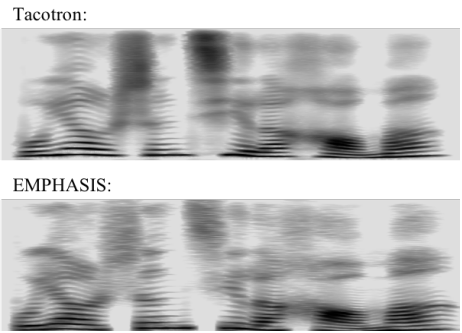

Figure 4: Spectrograms of segmentations of synthesized waveform by Tacotron and EMPHASIS

EMPHASIS yields slightly better preference score than Tacotron, the preferences of the subjects are not significant. The advantage of Tacotron is that it does not need phoneme segmentation, and the auto-alignment mechanism gives the speech more natural phoneme durations. However, its drawback is that the audio has fuzzy effect at high frequencies because of the use of Mel-scale spectrogram as the target. The spectral resolution at the high frequencies is relatively lower, and it is hard to restore the linear-scale spectrogram even with a post-processing network. We demonstrate this effect in Figure 4 by comparing two spectrograms. Using conditioning WaveNet [6] as wave generator may solve this problem, but there would be too much computational cost. EMPHASIS predicts linear-scale spectrogram directly, therefore, the audio quality is relatively higher. Griffin-Lim reconstruction algorithm is also a time-consuming method compared to WORLD, which makes Tacotron slower.

The concatenative system is inferior to EMPHASIS. Even though he HMM-based unit selection method has the best audio quality, it does not work well under this amount of training data, because there will be too much discontinuous points, especially when the training sentences have many emotional expressions.

## 4. Conclusions

The experimental results show that EMPHASIS is a successful TTS acoustic model. There are three key points for its success. Firstly, the feature grouping strategy in the input layer can extract emotional & prosodic features and phoneme-related features independently, which ensures that the emotion-related features are not ignored at the very beginning. Secondly, the CBHG-based network structure is powerful for extracting high-level representations from linguistic sequences. Finally, the independent bidirectional GRU layers reduce the interferences between different acoustic parameters, especially energy and LF0. EMPHASIS is designed to be a multi-lingual model, because the linguistic feature contains information from both Mandarin and English. It is a potential acoustic model for TTS system and can be improved in many ways, such as speaker-adaptive training, more complex affective speech generation (such as happy, anger and sad etc.) and synthesis of other languages, which will be our future works.

## 5. Acknowledgements

The authors would like to thank Feiya Li of Baidu Speech Department for organizing subjective evaluations.